\documentclass[reprint, aps, prl, floatfix,superscriptaddress]{revtex4-1}
\usepackage{graphicx}    
\usepackage{dcolumn}    
\usepackage{bm}             
\usepackage{amssymb}   
\usepackage{amsmath}    
\usepackage{amsthm}
\usepackage{mathrsfs}    
\usepackage{commath}   
\usepackage{subfigure}    
\usepackage{braket}
\usepackage{natbib}
\usepackage{float}
\usepackage{color}
\usepackage{siunitx}
\usepackage[colorlinks=true, allcolors=blue]{hyperref}
\usepackage{hyphenat}
\usepackage{footnote}
\usepackage{notes2bib}
\usepackage{xcolor}

\newcommand{\beq}{\begin{equation}}
\newcommand{\eeq}{\end{equation}}

\begin{document}

\title{Quantum transduction is enhanced by single mode squeezing operators}

\author{Changchun Zhong}
\email{zhong.changchun@uchicago.edu}
\affiliation{Pritzker School of Molecular Engineering, University of Chicago, Chicago, IL 60637, USA}

\author{Mingrui Xu}
\affiliation{Department of Electrical Engineering, Yale University, New Haven, CT 06520, USA}

\author{Aashish Clerk}
\affiliation{Pritzker School of Molecular Engineering, University of Chicago, Chicago, IL 60637, USA}

\author{Hong X. Tang}
\affiliation{Department of Electrical Engineering, Yale University, New Haven, CT 06520, USA}
\affiliation{Yale Quantum Institute, Yale University, New Haven, CT 06520, USA}

\author{Liang Jiang}
\email{liang.jiang@uchicago.edu}
\affiliation{Pritzker School of Molecular Engineering, University of Chicago, Chicago, IL 60637, USA}

\date{\today}

\begin{abstract}
Quantum transduction is an essential ingredient in scaling up distributed quantum architecture and is actively pursued based on various physical platforms. However, demonstrating a transducer with positive quantum capacity is still practically challenging. In this work, we discuss a new approach to relax the impedance matching condition to half impedance matching condition, achieved by introducing two-photon drive in the electro-optic transducer. We show the quantum transduction capacity can be enhanced and can be understood in a simple interference picture with the help of Bloch-Messiah decomposition. The parameter regimes with positive quantum capacity is identified and compared with and without the drive, indicating that the parametric drive-induced enhancement is promising in demonstrating quantum state conversion, and is expected to boost the performance of transduction with various physical platforms.

\end{abstract}

\maketitle

\textit{Introduction}.---Superconducting qubit based quantum networks---processing quantum information with superconducting circuits \cite{blais2021} and transmitting quantum signals by optical photons \cite{yin2017}---are appealing architectures for future communication systems \cite{Cirac1997,Kimble2008}. To scale up such networks, coherent conversion between microwave and optical (MO) states---quantum transduction---is indispensable as superconducting qubits lack intrinsic optical transitions. However, realizing MO quantum states conversion turns out to itself be extremely challenging with current technology \cite{andrews2014,vainsencher16,mirhosseini2020,zeuthen2020}. {Considering the transduction as a quantum channel, to obtain a positive quantum channel capacity}, the channel must have both high {channel transmissivity} and {low added noise}, e.g., at least $50\%$ transmissivity is needed \cite{weedbrook2021}. Although significant advances have been made recently \cite{Regal2011,Bochmann2013,Taylor2011,Barzanjeh2011,Wang2012,Tian2010,*Tian2012,*Tian2014,Midolo2018,Bagci2014,Winger2011,Pitanti2015,Tsang2010,*Tsang2011,Javerzac-Galy2016,Fan2018,fu2021,Hisatomi2016,zhu2020,han2018}, the traditional direct quantum transducer (DQT), which linearly converts MO quantum signals by beam splitter coupling, is still below the threshold where quantum capacity is zero.


A perfect transducer has infinite quantum capacity, which requires even more stringent impedance matching condition \cite{safavi2011}. {In recent years, respecting the fact that quantum capacity can be enhanced with two-way communication \cite{bennett1997,jeffrey2009}, there are approaches to use classical channel 
\cite{Barzanjeh2012,zhong2020,rueda2019,zhong2022,wu2021}, adaptive control \cite{Higginbotham2018,Meng18}, multi-pass interference \cite{lau2019}, etc.~to overcome the challenge of full impedance matching. Although encouraging, all these schemes require additional classical control, access to other output ports, or even infinitely squeezed input fields.}


For future high-bandwidth quantum transduction, it is still desirable to construct a DQT. In this paper, we discuss a {new DQT scheme based on a cavity electro-optic (EO) transducer \cite{sahu2022}. By applying a parametric drive to the microwave mode, we find a new approach that can relax the matching condition---two transduction quadratures are impedance matched---to \textit{half matching condition}---only one quadrature needs to be matched \cite{safavi2011,lau2019}. {We show that a generic transducer satisfying the half-matching condition yields an infinite quantum capacity. This capacity can be achieved using solely operations on the transudcer's active input and output modes, respecting the channel coding theorem \cite{wilde2013}. Note that the utility of half-impedance matched quantum transducers was discussed in Ref.~\cite{lau2019} in a slightly different context, and without any direct connection to quantum capacity. 
} 

Note this technique with {parametric drive has been used in optomechanical system for canceling counter-rotating term in transduction \cite{lau2020}, cooling \cite{sumei2009}, enhancing light-matter interaction \cite{wei2018,lu2015,leroux2018,peano2015,chen2022} and non-reciprocity \cite{tang2021}.} When there is {no intrinsic loss} and zero detuning, we show the parametric drive can enhance the transduction channel capacity in a broad parameter space. This observation is appealing especially given the current challenges in transducer design and fabrication to simultaneously achieve near-unity transmissivity and ground-state conversion.

{Parametric drive-enabled half-matching condition can also function even in the presence of intrinsic noise and finite detuning.} In general, the half matching condition can't be fully achieved by the drive, and the parametric-amplified noise could severely degrade the channel. {To keep the transduction capacity as large as possible, we need to maintain an enhanced channel transmissivity, and properly handle the parametric-amplified noise.} With the help of \textit{Bogoliubov transformation}, we show that near {the amplification threshold}, an exponential increase in the coupling strength is expected between the optical and microwave {Bogoliubov} mode. The parametric drive also induces amplified noise to the Bogoliubov mode, degrading the transduction performance. Nevertheless, we find the capacity still gains from the coupling enhancement in some parameter regime. More interestingly, the amplified noise can be eliminated completely if we implement the noise cancellation technique, e.g., generating broadband squeezed vacuum as the mode reservoir \cite{lemonde2016,tang2021}, thus the capacity enhancement can be maintained in larger parameter space.

\begin{figure}[t]
\centering
\includegraphics[width=\columnwidth]{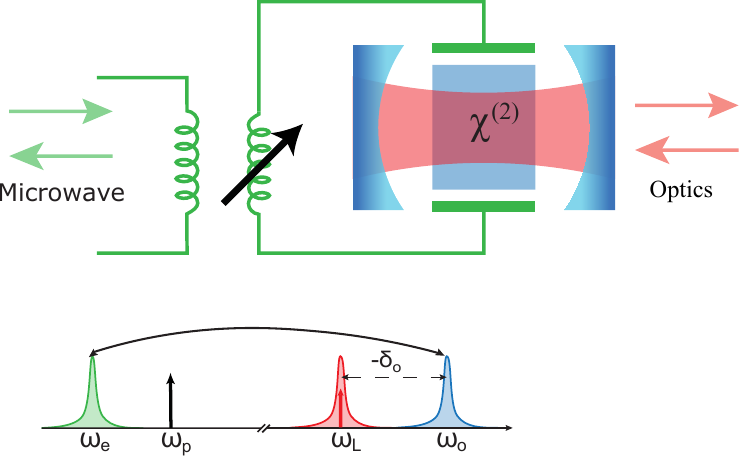}
\caption{Schematic figure for a cavity electro-optic transducer. An laser pump is applied on one of the optical modes (red) and it generates a beam splitter interaction between the optical $\hat{a}$ (blue) and microwave $\hat{b}$ (green), shown in the lower spectrum diagram. The microwave mode parametric squeezing is introduced through the tunable superconducting inductance, which is indicated by the black arrow. \label{fig1}}
\end{figure}

\textit{The model.}---Without losing generality, we take a cavity EO system for demonstration. Cavity EO is a hybrid superconducting-photonic device where a superconducting resonator is integrated with an optical cavity, consisting of material which features Pockels nonlinearity, e.g., AlN. As shown in Fig.~\ref{fig1}, the electrical field from the resonator can change the material refraction index, which modifies the frequency of optical photons. In reverse, microwave field can be modulated by the cavity optical fields due to the optical \textit{rectification} of Pockels material. With the material nonlinearity and a triple resonant design \cite{Fan2018}, a three wave mixing between two chosen optical modes and an electrical resonator can be realized, and by appropriately driving one of the two optical modes with laser frequency $\omega_\text{L}$, a beam splitter interaction can be generated $\hbar g(\hat{a}^\dagger\hat{b}+\hat{a}\hat{b}^\dagger)$, where $\hat{a}$ and $\hat{b}$ denote the optical and microwave mode operators, respectively. $g$ is the laser enhanced coupling strength. {With the system on resonance}, one can get a transduction channel---a single mode Bosonic loss channel with transmissivity (i.e. determinant of the quadrature-basis transmission matrix) \cite{chen2022,suppl}
\begin{equation}\label{eqtrans}
    \eta=\frac{4C_g}{(1+C_g)^2}\zeta_\text{e}\zeta_\text{o},
\end{equation}
where $C_g\equiv4g^2/\kappa_\text{o}\kappa_\text{e}$ is the MO \textit{cooperativity}. We have denoted ($\kappa_\text{o},\kappa_\text{e}$) as the (optical, microwave) total dissipation rates and ($\zeta_\text{o},\zeta_\text{e}$) as the (optical, microwave) extraction ratio \cite{exratio}. Currently, EO transducer's transmissivity is reported to be around several percent, which is still below the $50\%$ threshold for a general Bosonic loss channel to have positive quantum capacity \cite{sahu2022,suppl}. 


In this paper, we consider enhancing the transduction by introducing parametric drive to the superconducting resonator, {which can be realized in a superconducting resonator with intrinsic tunable inductance and is commonly used in cavity-based Josephson amplifiers \cite{malnou2018}.} The total Hamiltonian is given as
\begin{equation}\label{hamil}
    \begin{split}
        \hat{H}/\hbar=&-(\delta_\text{o}+\frac{\omega_\text{p}}{2})\hat{a}^\dagger\hat{a}+(\omega_\text{e}-\frac{\omega_\text{p}}{2})\hat{b}^\dagger\hat{b} \\&+g(\hat{a}^\dagger\hat{b}+\hat{a}\hat{b}^\dagger)
        +\nu(e^{-i\theta}\hat{b}^{\dagger^2}+e^{i\theta}\hat{b}^2),
    \end{split}
\end{equation}
where $\delta_\text{o}=\omega_\text{L}-\omega_\text{o}<0$ is the optical drive detuning. $\nu$, $\omega_\text{p}$ and $\theta$ are the parametric pump strength, frequency and phase, respectively. Note the Hamiltonian is written in the rotating frame of half of the parametric pump. Quite interestingly, when the system is on resonance, the channel transmissivity becomes \cite{suppl}
\begin{equation}
    \eta_\nu=\frac{4C_g}{(1+C_g)^2-4C_\nu}\zeta_\text{e}\zeta_\text{o},
\end{equation}
where we call $C_\nu\equiv{4\nu^2}/{\kappa_\text{e}^2}$ as the \textit{squeezing cooperativity}. In this paper, without state it otherwise, we are interested in the stable regimes when $C_\nu<(1+C_g)^2/4$.

In Fig.~\ref{fig2}, we numerically plot the consequences of introducing this drive while keeping other parameters ideal, e.g., unit extraction ratios and system on resonance. Figure~\ref{fig2}(a) shows the channel transmissivity with different squeezing as a function of the MO cooperativity. The parametric drive in general increases the transmissivity, indicating a positive capacity is potentially achievable with smaller $C_g$, thus {relaxing the demanding matching condition for the experiment}. Figure~\ref{fig2}(b) gives the quantum capacity lower bound of the transduction channel (see a brief review of quantum channel in the appendix \cite{suppl,eisert2005}). Indeed, we see that the parametric drive changes the channel behaviors: increases the quantum capacity for all $C_g$. The shining white curve is determined by
\begin{equation}\label{cond1}
    C_\nu=\frac{1}{4}(1-C_g)^2,
\end{equation}
and we will show later this curve exactly traces the parameters that make the transducer fulfill the \textit{half matching condition}, giving the transducer an optimal quantum capacity. Across the line the channel switches between a Bosonic loss channel and an amplification channel. Note the system is stable as long as the squeezing is not extremely high, indicated by the yellow dashed line in Fig.~\ref{fig2}(b). This squeezing enhanced quantum transduction channel is our first key observation in this paper.

\begin{figure*}[t]
\centering
\includegraphics[width=\textwidth]{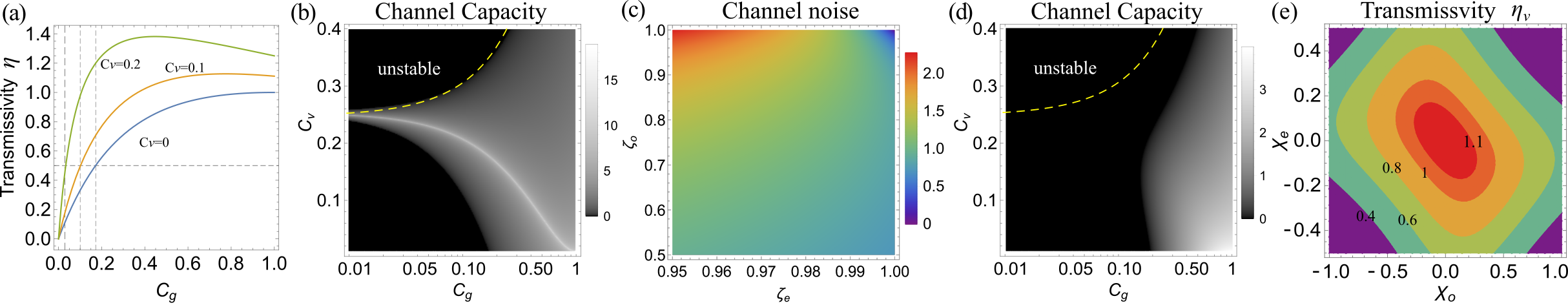}
\caption{(a) The channel transmissivity as a function of the MO cooperativity $C_g$ with squeezing cooperativities $C_\nu=0,0.1,0.2$, respectively. The horizontal dashed lines mark the boundary where the system reaches the positive quantum capacity threshold. The vertical lines help in viewing the corresponding $C_g$ threshold. (b)The quantum capacity lower bound in terms of $C_\nu$ and $C_g$. The eye-catching white line marks the boundary where the channel transmissivity crosses unity---the system goes from a Bosonic loss channel to an amplification channel. In (a) and (b), the extraction ratios are taken to be perfect in order to highlight the key consequences of introducing intracavity squeezing. (c) The channel noise as a function of extraction ratios, with $C_g=0.14,C_\nu=0.16$. (d) The capacity lower bound as in (b) with extraction ratios $\zeta_\mathrm{o}=0.95,\zeta_\mathrm{e}=0.99$. (e) The channel transmissivity with respect to the detuning, with $C_g=0.4,C_\nu=0.15,\zeta_\mathrm{e}=0.99,\zeta_\mathrm{o}=0.95$. \label{fig2}}
\end{figure*}

\textit{Euler decomposition of scattering matrix}---A clear physical picture of the enhancement can be obtained by looking at the scattering matrix structure. {For system on resonance and with unit extraction ratios}, the scattering matrix $\mathbf{S}_\mathrm{x}$ which {connects the input and output mode quadratures $\mathbf{x}_\mathrm{in}=(\hat{x}^a_\mathrm{in},\hat{p}^a_\mathrm{in},\hat{x}^b_\mathrm{in},\hat{p}^b_\mathrm{in})^\mathrm{T}$, $\mathbf{x}_\mathrm{out}=(\hat{x}^a_\mathrm{out},\hat{p}^a_\mathrm{out},\hat{x}^b_\mathrm{out},\hat{p}^b_\mathrm{out})^\mathrm{T}$} is given by a $4\times 4$ symplectic matrix \cite{suppl}. For any symplectic matrix, we can always find a Bloch-Messiah decomposition $\mathbf{S}_\mathrm{x}=\mathbf{ODO^\prime}$ \cite{gosson2006}, where $\mathbf{O,O^\prime}$ are symplectic orthogonal matrices---characterizing beam splitters and phase shifters---and $\mathbf{D}$ is a diagonal matrix---characterizing squeezing of individual mode (see Ref.~\cite{suppl} for expressions). Remarkably, {the parameter $C_\nu$ only appears in $\mathbf{D}$ matrix, which means the scattering process can be characterized as a local squeezer sandwiched by two $C_\nu$-independent transformations.} Such kind of transduction channel can in general made to satisfy the \textit{half match condition} by tuning the squeezer in the middle. To get a flavor of the physics, we write down {the quadrature transformation from $\mathbf{S_x}$}
\begin{equation}
    \begin{split}
        \hat{x}^b_\mathrm{out}=&\frac{\sqrt{C_g}(1+r)}{1+C_g}\hat{p}^a_\mathrm{in}+\frac{{r-C_g}}{(1+C_g)}\hat{x}^b_\mathrm{in},\\
        \hat{p}^b_\mathrm{out}=&-\frac{\sqrt{C_g}(1+{1/r})}{1+C_g}\hat{x}^a_\mathrm{in}+\frac{1/r-C_g}{1+C_g}\hat{p}^b_\mathrm{in},
    \end{split}
\end{equation}
where $r=\frac{1+C_g+2\sqrt{C_\nu}}{1+C_g-2\sqrt{C_\nu}}$. Obviously, if $r=C_g$ (leading to Eq.~\ref{cond1}), we have $\hat{x}^b_\mathrm{out}=\sqrt{C_g}\hat{p}^a_\mathrm{in},\hat{p}^b_\mathrm{out}=-1/\sqrt{C_g}\hat{x}^a_\mathrm{in}+(1/C_g-1)\hat{p}^b_\mathrm{in}$ where the reflection in $\hat{x}^b_\mathrm{out}$ quadrature is completely canceled by interference. We thus achieved the half matching condition that one quadrature is reflectionless and the other quadrature is not. As we show, a transduction channel with \text{half matching condition} has infinite quantum capacity, which should be called perfect transduction. Releasing the perfect matching condition to \textit{half matching condition} as perfect transduction channel is our another key contribution and shall be significantly helpful in future transducer designs \cite{safavi2011,lau2019}. 


\textit{Bogoliubov mode and {near amplification threshold} transduction.}---The above discussion captures the major physics with unit extraction ratios $\zeta_\mathrm{e,o}=1$. If the extraction ratios are not unity, {the system couples to extra source of noise which will be amplified and contaminate the output signal. This would potentially kill the gain of introducing the parametric drive.} In general, the channel noise expression has a very complicated dependence on ($C_g,C_\nu,\zeta_\mathrm{e},\zeta_\mathrm{o}$) and we numerically show it in Fig.~\ref{fig2}(c). The channel noise quickly increases as the extraction ratios deviates from one, which severely degrade the quantum capacity, as shown in Fig.~\ref{fig2}(d). Handling the amplified noise is thus essential in maintaining the enhancement. 

In practice, besides adding squeezing, we can play with another knob---the system detuning---to change the channel transmissivity \cite{suppl}. In general, we have
\begin{equation}
    \eta_\nu=\frac{4C_g\zeta_\mathrm{o}\zeta_\mathrm{e}}{ C_g^2+C_g(2+8\chi_\mathrm{e}\chi_\mathrm{o})+(1-4C_\nu+4\chi_\mathrm{e}^2)(1+4\chi_\mathrm{o}^2) },
\end{equation}
where $\chi_\mathrm{o}={\Delta_\mathrm{o}}/{\kappa_\mathrm{o}}$, $\chi_\mathrm{e}={\Delta_\mathrm{e}}/{\kappa_\mathrm{e}}$ and $\Delta_\text{o}\equiv\delta_\text{o}+\omega_\text{p}/2$, $\Delta_\text{e}\equiv\omega_\text{e}-\omega_\text{p}/2$. As shown in Fig.~\ref{fig2}(e), the channel transmissivity decreases as we increase the system detuning, which is within expectation for a less resonant system. Note reducing $\eta_\nu$ can also simply achieved by lowering the squeezing. {However, for large detuned regime, we can use the tool of \textit{Bogoliubov} transformation to better understand the enhancement, and noise elimination can be performed such that capacity degradation is mitigated.} 

\color{black}

\begin{figure*}[t]
\centering
\includegraphics[width=\textwidth]{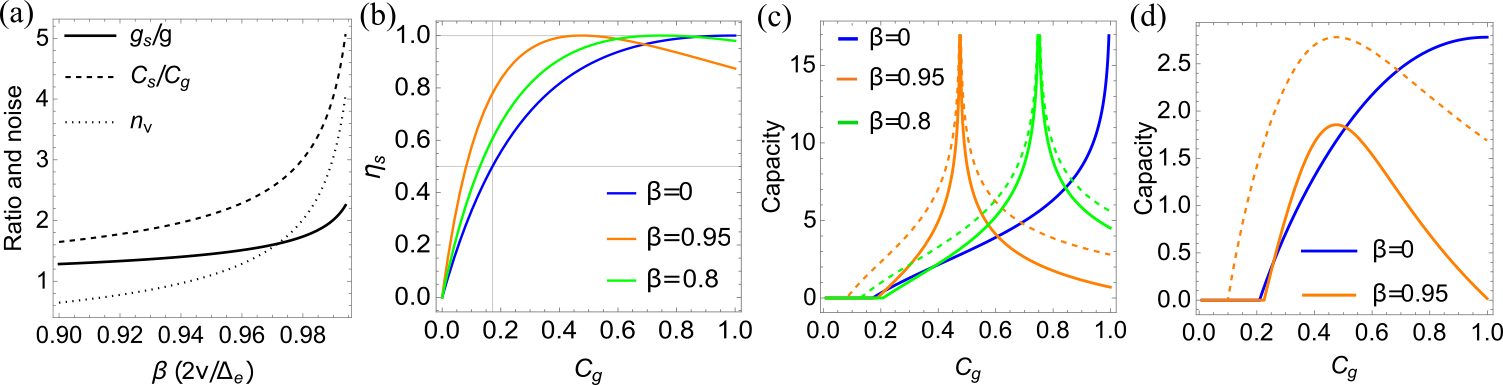}
\caption{(a) The coupling strength and cooperativity enhancement, and the squeezing induced noise in terms of the parameter $\beta=2\nu/\Delta_\text{e}$. (b) and (c) depict the transmissivity $\eta_\text{s}$ and the quantum capacity lower bound as a function of the MO cooperativity with giving $\beta=\{0,0.8,0.95\}$, corresponding to the solid blue, green and orange curves. The solid curves in (c) show the lower bound including squeezing induced noise, and the dashed orange and green lines in (c) show the same bound by assuming the induced noise is eliminated. Extraction ratios are assumed to be one in (b)(c). (d) The capacity lower bound with respect to $C_g$ for $\beta=\{0,0.95\}$ and extraction ratios $\zeta_\mathrm{e}=0.97,\zeta_\mathrm{o}=0.9$. The orange solid (dashed) includes (excludes) the squeezing induced noises. Note the noise elimination needs an input squeezing that is equal to the squeezing from the parametric drive, which is $10\log_{10}e^{4\nu}$ dB. \label{fig3}}
\end{figure*}

Define a Bogoliubov $\hat{b}_\text{s}$ through the transformation \cite{scully1997,agarwal2012} $\hat{b}_\text{s}=\cosh(r)\hat{b}+e^{-i\theta}\sinh(r)\hat{b}^\dagger$, where the mode dissipation rate $\kappa_\text{s}=\kappa_\text{e}$, and the parameter $r$ is the \textit{effective squeezing} which satisfies $\tanh(2r)=2\nu/\Delta_\text{e}\equiv\beta$. With the Bogoliubov transformation, we can rewrite the Hamiltonian Eq.~\ref{hamil} in terms of the Bogoliubov mode as
\begin{equation}\label{hamils}
    \begin{split}
        \hat{H}_\text{s}/\hbar=&-\Delta_\text{o}\hat{a}^\dagger\hat{a}+\omega_\text{s}\hat{b}_\text{s}^\dagger\hat{b}_\text{s}+g_\text{s}(\hat{a}^\dagger\hat{b}_\text{s}+\hat{a}\hat{b}^\dagger_\text{s}),
    \end{split}
\end{equation}
where $\omega_\text{s}=\sqrt{\Delta_\text{e}^2-(2\nu)^2}$. Equation~\ref{hamils} takes exactly the same form as a beam splitter interaction, except that the coupling strength $g$ is replaced by $g_\text{s}=g\cosh(r)$ which can be exponentially enhanced by the effective squeezing. Obviously, the channel transmissivity is obtained as
\begin{equation}
    \eta_\mathrm{s}=\frac{4C_\text{s}}{(1+C_\text{s})^2}\zeta_\text{o}\zeta_\text{s},
\end{equation}
where $C_\text{s}=4g_\text{s}^2/\kappa_\text{o}\kappa_\text{s}$ is Bogoliubov-optical (BO) mode cooperativity, and $\zeta_\text{s}$ is the Bogoliubov mode extraction ratio. Note this transmissivity $\eta_\mathrm{s}$ is defined in terms of the Bogoliubov mode which shall not be confused with $\eta_\nu$. Since the BO cooperativity depends on the square of the coupling strength, the channel transmissivity, which depends on the cooperativity, is expected to increase as the coupling strength is enhanced. As shown in Fig.~\ref{fig3}(a), the solid and dashed curves depict the exponential enhancement of the coupling strength and cooperativity as {the parameter $\beta$} approaches one, which corresponds {the \textit{amplification threshold} of the system} \cite{Aspelmeyer2014}. The change of channel transmissivity $\eta_\mathrm{s}$ is shown in Fig.~\ref{fig3}(b) with respect to the MO cooperativity. We see it is possible by adding squeezing to get higher transmissivity even if $C_\text{g}$ is small. The orange curve ($\beta=0.95$) can get to the $\eta_\mathrm{s}=0.5$ threshold with much smaller $C_g$ than the blue curve (no squeezing). 


However, this coupling enhancement comes with a side effect---the squeezing-amplified noise which couples to the Bogoliubov mode. This noise is given by $n_\nu=\cosh(2r)n_\text{th}+\sinh^2(r)$, where $n_\mathrm{th}$ is the thermal photon of microwave bath \cite{suppl}. In the ideal case, the microwave resonator sits in a very cold environment, and without state it otherwise we assume it to be vacuum in the numerical evaluations. Still, the noise $n_\nu$ is proportional to $\sinh^2(r)$, which could possibly ruin the channel from realizing any quantum conversion as the system gets close to {amplification threshold}. As shown by the dotted line in Fig.~\ref{fig3}(a), the squeezing-amplified noise also increases along with $\beta$. 

Admittedly, the coupling enhancement and the amplified noise are a pair of competing factors. Fortunately for quantum transduction, we still gain from the transmissivity enhancement in some regime and we can benefit even more if we adopt the technique of noise elimination (see \cite{suppl} for details of noise elimination). The evidence is given by the quantum channel capacity lower bound, as shown in Fig.~\ref{fig3}(c) for unit extraction ratios and Fig.~\ref{fig3}(d) for $\zeta_\mathrm{e}=0.97$ and $\zeta_\mathrm{o}=0.9$. In Fig.~\ref{fig3}(c), the blue curve shows the lower bound for $\beta=0$. As we increase $\beta$, the optimal capacity can be achieved for $C_g$ being smaller, as shown by the green and orange solid curves. Moreover, if the technique of noise elimination is adopted \cite{suppl,lemonde2016,tang2021}, the capacity gets even bigger and the positive value spread to larger parameter regime, as depicted by the dashed green and orange curves. Figure~\ref{fig3}(d) gives the capacity lower bound degraded by the non-unity extraction ratios. Still, larger capacity can be achieved by the parametric drive, and noise cancellation further enhances the transduction channel in broader parameter regimes, as shown by the orange solid and dashed curves. Since designing a transducer with large $C_g$ is hard and severely hinders the experimental realization, the observation of smaller $C_g$ for transduction described in this section is expected to accelerate the pace in ultimately realizing the MO quantum state conversion. 


\textit{Discussion.}---{A general transducer with half matching condition is given by
\begin{equation}
    \begin{split}
        \begin{cases}
         \hat{x}^b_\mathrm{out}=\xi\hat{x}^a_\mathrm{in}+\gamma\hat{x}^b_\mathrm{in}\\
         \hat{p}^b_\mathrm{out}=\frac{1}{\xi}\hat{p}^a_\mathrm{in} 
        \end{cases}
        \begin{cases}
        \hat{x}^a_\mathrm{out}=\frac{1}{\xi}\hat{x}^b_\mathrm{in}\\
        \hat{p}^a_\mathrm{out}=\xi\hat{p}^b_\mathrm{in}-\gamma\hat{p}^a_\mathrm{in}
        \end{cases}
    \end{split}.
\end{equation}
Take $a\rightarrow b$ for example, one can show perfect transduction can be realized by first squeezing (anti-squeezing) the input $\hat{p}_\mathrm{in}^a$ ($\hat{x}_\mathrm{in}^a$) then squeezing (anti-squeezing) the output $\hat{x}_\mathrm{out}^b$ ($\hat{p}_\mathrm{out}^b$) without resorting to any control on other inputs or outputs. Note this differs from the main approach suggested in Ref.~\cite{lau2019}. In addition, we point out that the half-matched transducer defines a two-way perfect transduction channel. To achieve that, one can do encoding by squeezing the input quadratures ($\hat{x}_\mathrm{in}^b,\hat{p}_\mathrm{in}^a$) while anti-squeezing ($\hat{p}_\mathrm{in}^b,\hat{x}_\mathrm{in}^a$), and do decoding at the output by squeezing ($\hat{x}^b_\mathrm{out},\hat{p}^a_\mathrm{out}$) and anti-squeezing ($\hat{x}^a_\mathrm{out},\hat{p}^b_\mathrm{out}$).}

The rotating wave approximation in Eq.~\ref{hamils} requires the weak coupling $\omega_\mathrm{s}>g_\mathrm{s}$ and $\abs{\Delta_\mathrm{o}}+\omega_\mathrm{s}\gg\abs{\Delta_\mathrm{o}}-\omega_\mathrm{s}$, indicating $\omega_\mathrm{s}\neq 0$, thus forbids $\beta$ becoming one. The Bogoliubov transformation also requires $\beta<1$ which means the system can't really approach the {amplification threshold}. Fortunately, as shown by the orange curve in Fig.~\ref{fig3} with $\beta=0.95,0.8$, the channel enhancement is already enormous when the system is close to the threshold, thus we give it the name ``near amplification threshold enhancement". 
{Note the constraint $\beta<1$ is purely from the Bogoliubov framework. Practically, $\beta$ can be any physical value and how to remove the corresponding squeezing-amplified noise is an interesting topic} for the future.

A possible extension of the scheme is to further introduce two photon drive on the optical mode and explore how the half matching condition develops. In the Bogoliubov picture, this could also strengthen the coupling strength. It is worth mentioning that as the coupling strength increases (larger than the mode loss), one might enter the strong coupling regime where mode splitting will happen in the conversion spectrum \cite{zhong2020_}. In this case, the optimal transmissivity is not at $\omega=0$ any more (currently we assume the system is weakly coupled) and should be taken into account in future implementations.

\begin{acknowledgments}
We acknowledge support from the ARO (W911NF-18-1-0020, W911NF-18-1-0212), ARO MURI (W911NF-16-1-0349, W911NF-21-1-0325), AFOSR MURI (FA9550-19-1-0399, FA9550-21-1-0209), AFRL (FA8649-21-P-0781), DOE Q-NEXT, NSF (OMA-1936118, EEC-1941583, OMA-2137642), NTT Research, and the Packard Foundation (2020-71479).
\end{acknowledgments}

\bibliographystyle{apsrev4-1}
\bibliography{er}

\begin{thebibliography}{61}
\bibitem [{exr()}]{exratio}%
  \BibitemOpen
  \href@noop {} {}\bibinfo {note} {The extraction ratio is defined as signal
  coupling rate to the total loss rate, e.g.,
  $\zeta_\text{o}\equiv{\kappa_\text{o,c}}/{\kappa_\text{o}}$, where
  $\kappa_\text{o,c}$ is the signal coupling rate, and
  $\kappa_\mathrm{o,i}=\kappa_\mathrm{o}-\kappa_\mathrm{o,c}$ is called
  intrinsic loss rate.}\BibitemShut {Stop}%
\end{thebibliography}

\onecolumngrid

\section{\textit{Supplemental material for} ``{Quantum transduction is enhanced by single mode squeezing operators}"}

\section{A---Gaussian quantum channel}

A Gaussian quantum channel $\mathcal{N}:\rho_\text{in}\rightarrow\rho_\text{out}$ can be specified by its action on the statistical first and second moments of arbitrary Gaussian state $\hat{\rho}(\bar{\mathbf{x}},\mathbf{V})$. In general, we have
\begin{equation}
\begin{split}
    \bar{\mathbf{x}}&\rightarrow\mathbf{T\bar{x}+d},\\
    \mathbf{V}&\rightarrow\mathbf{TVT}^T+\mathbf{N},
\end{split}
\end{equation}
where $\mathbf{T,N}$ are real matrices satisfying the channel completely positive condition $\mathbf{N+i\Omega-iT\Omega T^T}\ge 0$. Specifically, when $\mathbf{N}=\mathbf{0}$, $\mathbf{T}$ is a symplectic matrix, which defines a Gaussian unitary channel. Given a single mode Gaussian channel, the quantum capacity is lower bounded by the following expression
\begin{equation}
Q_\text{LB}=
\begin{cases}
      \max\{0,\log_2|\frac{\eta}{1-\eta}|-g(n_e)\}, & \eta\neq 1 \\
      \max\{0,\log_2(\frac{2}{e\sigma^2})\},& \eta=1, 
\end{cases}
\end{equation}
where $g(n_e)=(n_e+1)\log_2(n_e+1)-n_e\log_2n_e$. $\eta$ and $n_e$ are the channel transmissivity and added noise, respectively, which are given by
\begin{equation}
    \begin{split}
        \eta&=\det\mathbf{T},\\
        n_e&=
        \begin{cases}
         \frac{\sqrt{\det\mathbf{N}}}{2\abs{1-\eta}}-\frac{1}{2}, & \eta\neq 1\\
         \sqrt{\det\mathbf{N}}=\sigma^2,& \eta=1 
        \end{cases}.
    \end{split}
\end{equation}
{Note in some literature, the transmissivity is defined as the quadrature conversion efficiency, which can be varied for different quadratures for none isotropic channel.} Clearly, $\eta=1/2$ is the least threshold for a Bosonic loss channel to have non-zero quantum capacity.

\section{B---The transduction channel with squeezing}

In this section, we will derive the Gaussian transduction channel given by the cavity electro-optical system with microwave squeezing. As given in the main text, the Hamiltonian takes the form 
\begin{equation}\label{hamil2}
    \begin{split}
        \hat{H}/\hbar=-\Delta_\text{o}\hat{a}^\dagger\hat{a}+\Delta_\text{e}\hat{b}^\dagger\hat{b}+ g(\hat{a}^\dagger\hat{b}+\hat{a}\hat{b}^\dagger)
        +\nu(e^{-i\theta}\hat{b}^{\dagger^2}+e^{i\theta}\hat{b}^2).
    \end{split}
\end{equation}
We write down the Heisenberg-Langevin equations and input--output relations
\begin{equation} \label{se1}
\begin{split}
\dot{\mathbf{a}} &= \mathbf{A} \, \mathbf{a} + \mathbf{B} \, \mathbf{a}_\mathrm{in},\\
\mathbf{a}_{\mathrm{out}} &= \mathbf{B}^\mathrm{T} \mathbf{a} - \mathbf{a}_{\mathrm{in}},
\end{split}
\end{equation}
in which $\mathbf{a}$ is a vector which collects all the mode operators and similarly $\mathbf{a}_\mathrm{in}$ and $\mathbf{a}_\mathrm{out}$ collect all the input and output mode operators
\begin{equation}
\begin{split}
\mathbf{a} &= \left( \hat{a}, \, \hat{a}^\dagger, \, \hat{b}, \, \hat{b}^\dagger \right)^\mathrm{T}\\
\mathbf{a}_{\mathrm{in}} &= \left(\hat{a}_{\mathrm{in,c}}, \, \hat{a}^\dagger_{\mathrm{in,c}}, \, \hat{a}_{\mathrm{in,i}}, \, \hat{a}^\dagger_{\mathrm{in,i}}, \, \hat{b}_{\mathrm{in,c}}, \, \hat{b}^\dagger_{\mathrm{in,c}}, \, \hat{b}_{\mathrm{in,i}}, \, \hat{b}^\dagger_{\mathrm{in,i}}\right)^\mathrm{T} \\
\mathbf{a}_\mathrm{out} &= \left(\hat{a}_\mathrm{out,c}, \, \hat{a}^\dagger_\mathrm{out,c}, \, \hat{a}_\mathrm{out,i}, \, \hat{a}^\dagger_\mathrm{out,i}, \, \hat{b}_\mathrm{out,c}, \, \hat{b}^\dagger_\mathrm{out,c}, \, \hat{b}_{\mathrm{out,i}}, \, \hat{b}^\dagger_{\mathrm{out,i}}\right)^\mathrm{T}\\
\mathbf{A} &=
\begin{pmatrix}
i\Delta_\mathrm{o}-\frac{\kappa_{\mathrm{o}}}{2}	&	 0	&	-ig	&	0	\\
0	&	-i\Delta_\mathrm{o}-\frac{\kappa_{\mathrm{o}}}{2}	&	0	&	ig	 \\	
-ig	&	0	&	-i\Delta_\mathrm{e}-\frac{\kappa_{\mathrm{e}}}{2}	&	-i2\nu e^{-i\theta}	 \\
0	&	ig	&	i2\nu e^{i\theta}	&	i\Delta_\mathrm{e}-\frac{\kappa_{\mathrm{e}}}{2}	 \\
\end{pmatrix}, \\
\mathbf{B} &=
\begin{pmatrix}
\sqrt{\kappa_{\mathrm{o,c}}}	&	 0	&	\sqrt{\kappa_{\mathrm{o,i}}}	&	0	&	0	&	0	&	0	&	0 \\
0	&	 \sqrt{\kappa_{\mathrm{o,c}}}	&	0	&	\sqrt{\kappa_{\mathrm{o,i}}}	&	0	&	0	&	0	&	0  \\
0	&	 0	&	0	&	0	&	\sqrt{\kappa_{\mathrm{e,c}}}	&	0	&	\sqrt{\kappa_{\mathrm{e,i}}}	&	0 \\
0	&	 0	&	0	&	0	&	0	&	\sqrt{\kappa_{\mathrm{e,c}}}	&	0	&	\sqrt{\kappa_{\mathrm{e,i}}}  \\
\end{pmatrix}.
\end{split}
\end{equation}
The lower indices ``in,out" denote the ``input,output" modes, and the indices ``c,i" denote the ``coupling,intrinsic" loss ports. Transform all the mode operators into the frequency domain according to the formula,
\begin{align}
\mathbf{O}[\omega] &= \int^\infty_{-\infty} \mathbf{O}(t) e^{i\omega t} \: \mathrm{d}t.
\end{align}
Straightforwardly, we obtain that $\mathbf{a}_\mathrm{in}[\omega]$ and $\mathbf{a}_\mathrm{out}[\omega]$ is linked through a scattering matrix $\mathbf{S}_a[\omega]$,
\beq
\mathbf{a}_\mathrm{out}[\omega] = \mathbf{S}_a[\omega] \, \mathbf{a}_\mathrm{in}[\omega] = \left[ \mathbf{B}^\mathrm{T} \left(-i \omega \mathbf{D}_4-\mathbf{A} \right)^{-1} \mathbf{B} - \mathbf{I}_{8} \right] \mathbf{a}_\mathrm{in}[\omega],
\eeq
where $\mathbf{I}_{8}$ denotes the 8-dimensional identity matrix, and $\mathbf{D}_4=\mathrm{diag}(1, \, -1,\,1,\,-1)$. The scattering matrix can be rewritten in the quadrature basis using the expression
\begin{equation}
\begin{pmatrix}
\hat{x} \\ \hat{p}
\end{pmatrix}
=
\begin{pmatrix}
1&1\\
-i&i
\end{pmatrix}
\begin{pmatrix}
\hat{a}\\
\hat{a}^\dagger
\end{pmatrix}.
\end{equation}
We thus have a new scattering matrix defined in the quadrature basis
\beq\label{s8}
\mathbf{x}_\mathrm{out}[\omega] = \mathbf{S}_x[\omega] \, \mathbf{x}_\mathrm{in}[\omega] = \mathbf{Q} \, \mathbf{S}_a[\omega] \, \mathbf{Q}^{-1} \mathbf{x}_\mathrm{in}[\omega],
\eeq
where
\beq
\mathbf{Q}=\mathbf{I}_{4}\otimes\begin{pmatrix}
1&1\\
-i&i
\end{pmatrix}.
\eeq
The input and output quadrature fields are
\beq
\mathbf{x}_{\mathrm{in}} = \left(\hat{x}^a_{\mathrm{in,c}}, \, \hat{p}^a_{\mathrm{in,c}}, \, \hat{x}^a_{\mathrm{in,i}}, \, \hat{p}^a_{\mathrm{in,i}}, \, \hat{x}^b_{\mathrm{in,c}}, \, \hat{p}^b_{\mathrm{in,c}}, \, \hat{x}^b_{\mathrm{in,i}}, \, \hat{p}^b_{\mathrm{in,i}}\right)^\mathrm{T},
\eeq 
and 
\beq
\mathbf{x}_{\mathrm{out}} = \left(\hat{x}^a_\mathrm{out,c}, \, \hat{p}^a_\mathrm{out,c}, \, \hat{x}^a_\mathrm{out,i}, \, \hat{p}^a_\mathrm{out,i}, \, \hat{x}^b_\mathrm{out,c}, \, \hat{p}^b_\mathrm{out,c}, \, \hat{x}^b_{\mathrm{out,i}}, \, \hat{p}^b_{\mathrm{out,i}}\right)^\mathrm{T}.
\eeq
With the scattering matrix, we easily get the transduction channel. For example, from optical to microwave conversion, we have
\begin{equation}
    \begin{pmatrix}
    \hat{x}^b_\mathrm{out,c}\\
    \hat{p}^b_\mathrm{out,c}
    \end{pmatrix}=
    \mathbf{S}_x[\omega]^{\braket{5::6,1::2}}
    \begin{pmatrix}
    \hat{x}^a_\mathrm{in,c}\\
    \hat{p}^a_\mathrm{in,c}
    \end{pmatrix}+\mathbf{S}_x[\omega]^{\braket{5::6,3::8}}\cdot\mathbf{x}_\mathrm{in}^{\braket{3::8}},
\end{equation}
where the upper index $\braket{i::j,m::n}$ means the matrix elements from row $i$ to $j$ and column $m$ to $n$. Obviously, the covariance matrix is transformed according to 
\begin{equation}
    \mathbf{V}^b_\mathrm{out}=\mathbf{T}\mathbf{V}^a_\mathrm{in}\mathbf{T}^\mathrm{T}+\mathbf{N},
\end{equation}
which is exactly the form of a single mode Gaussian channel. Here we define the matrices
\begin{equation}
\begin{split}
    \mathbf{T}=&\mathbf{S}_x[\omega]^{\braket{5::6,1::2}}\\
    \mathbf{N}=&\mathbf{S}_x[\omega]^{\braket{5::6,3::8}}\braket{\mathbf{x}_\mathrm{in}^{\braket{3::8}}\cdot \mathbf{x}_\mathrm{in}^{{\braket{3::8}}^\mathrm{T} }} \mathbf{S}_x[\omega]^{{\braket{5::6,3::8}}^\mathrm{T}}. 
\end{split}
\end{equation}
Note the channel transmissivity is obtained as (when on resonance and $\omega=0$ in rotating frame)
\begin{equation}
    \eta_\nu=\det\mathbf{T}=\begin{cases}
     \frac{4C_g}{(1+C_g)^2-4C_\nu}\zeta_\mathrm{o}\zeta_\mathrm{e}, & \nu\neq 0 \\
     \frac{4C_g}{(1+C_g)^2}\zeta_\mathrm{o}\zeta_\mathrm{e}, & \nu = 0
    \end{cases}
\end{equation}
which is given in the main text. Note the transduction process is reciprocal, meaning the microwave to optical conversion has the same channel parameters.

Another interesting quantify is the transduction bandwidth as a function of the squeezing. For on resonance detuning, we have
\begin{equation}
    \eta_\nu[\omega]=\frac{4C_g\kappa_\mathrm{e}^2\kappa_\mathrm{o}^2\zeta_\mathrm{e}\zeta_\mathrm{o}}{[(1+C_g)^2-4C_\nu]\kappa_\mathrm{e}\kappa_\mathrm{o}+4[(1-4C_\nu)\kappa_\mathrm{e}^2-2C_g\kappa_\mathrm{e}\kappa_\mathrm{o}+\kappa_\mathrm{o}^2]\omega^2+16\omega^4 }.
\end{equation}
The numerical plot is shown in Fig.~\ref{appfig1}. We see the squeezing enhances the transmissivity in general, while the transduction bandwidth decreases.

\begin{figure}[t]
\centering
\includegraphics[width=8cm]{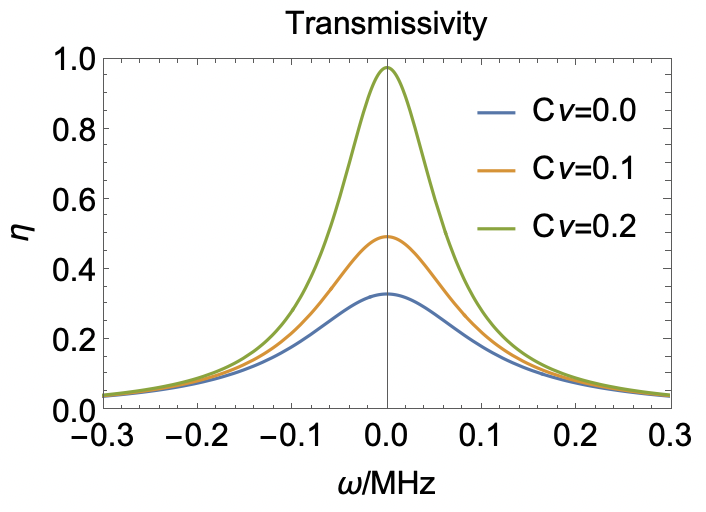}
\caption{The transmissivity as a function of the frequency for different squeezing parameters. The transduction bandwidth decreases as we increase the squeezing. The parameter $C_g=0.1,\zeta_e=\zeta_o=1$, $\kappa_\mathrm{o}=100$ MHz, $\kappa_\mathrm{e}=0.2$ MHz are used. \label{appfig1}}
\end{figure}

\section{C---The Euler decomposition}

In specific, the scattering matrix defined in Eq.~\ref{s8} can be reduced to a $4\times 4$ matrix when we take a unit extraction ratio
\begin{equation}
\mathbf{S}_x=
\begin{pmatrix}
\frac{1-C_g+2\sqrt{C_\nu}}{1+C_g+2\sqrt{C_\nu}} & 0 & 0 & \frac{2\sqrt{C_g}}{1+C_g+2\sqrt{C_\nu}} \\
0 & \frac{1-C_g-2\sqrt{C_\nu}}{1+C_g-2\sqrt{C_\nu}} & \frac{-2\sqrt{C_g}}{1+C_g-2\sqrt{C_\nu}} & 0 \\
0 & \frac{2\sqrt{C_g}}{1+C_g-2\sqrt{C_\nu}} & \frac{1-C_g+2\sqrt{C_\nu}}{1+C_g-2\sqrt{C_\nu}} & 0 \\
\frac{-2\sqrt{C_\nu}}{1+C_g+2\sqrt{C_\nu}} & 0 & 0 & \frac{1-C_g-2\sqrt{C_\nu}}{1+C_g+2\sqrt{C_\nu}}
\end{pmatrix},
\end{equation}
which connects the input and output quadratures of the coupling loss ports $\mathbf{x}_\mathrm{in,c}=(\hat{x}^a_\mathrm{in,c},\hat{p}^a_\mathrm{in,c},\hat{x}^b_\mathrm{in,c},\hat{p}^b_\mathrm{in,c})^\mathrm{T}$, $\mathbf{x}_\mathrm{out,c}=(\hat{x}^a_\mathrm{out,c},\hat{p}^a_\mathrm{out,c},\hat{x}^b_\mathrm{out,c},\hat{p}^b_\mathrm{out,c})^\mathrm{T}$. In the main text, we removed all the port index for simplicity. As classified in Ref.~\cite{lau2019}, this channel belongs to the class $[2,2]$, which is not perfect in general. If we have the squeezing parameter $C_\nu=\frac{1}{4}(1-C_g)^2$, the scattering matrix becomes
\begin{equation}
    \mathbf{S}_x=
\begin{pmatrix}
1-C_g & 0 & 0 & \sqrt{C_g} \\
0 & 0 & -\frac{1}{\sqrt{C_g}} & 0 \\
0 & \frac{1}{\sqrt{C_g}} & -1+\frac{1}{\sqrt{C_g}} & 0\\
-\sqrt{C_g} & 0 & 0 & 0
\end{pmatrix},
\end{equation}
which corresponds to the transducer {class $[2,1]$}, and can be made to be perfect by squeezing or Homodyne measurement. In the quantum channel language, the quantum capacity diverges. To understand why this is happening, we can Euler decompose the scattering matrix. In principle, any symplectic matrix $\mathbf{S}\in\mathrm{Sp}(2n,R)$ can be decomposed as the product of three matrices
\begin{equation}
    \mathbf{S=ODO}^\prime
\end{equation}
where $\mathbf{O},\mathbf{O}^\prime\in\mathrm{Sp}(2n,R)\cap O(2n)$ are symplectic orthogonal, and $\mathbf{D}$ is a diagonal matrix with local mode squeezing. This decomposition is called symplectic \textit{Euler decomposition} or \textit{Bloch-Messiah decomposition}. Note the matrix $\mathbf{D}$ only needs local squeezer, and the symplectic orthogonal matrix corresponds to those passive unitary transformations, which can be simply realized by beam splitter and phase shifter. In our case, we have the following clean form of decomposition
\begin{equation}
    \mathbf{S}_x=
    \begin{pmatrix}
    0 & \frac{1}{-\sqrt{1+C_g}} & 0 & \frac{\sqrt{C_g}}{\sqrt{1+C_g}} \\
    \frac{1}{\sqrt{1+C_g}} & 0 & \frac{-\sqrt{C_g}}{\sqrt{1+C_g}} & 0 \\
    \frac{\sqrt{C_g}}{\sqrt{1+C_g}} & 0 & \frac{1}{\sqrt{1+C_g}} & 0 \\
    0 & \frac{\sqrt{C_g}}{\sqrt{1+C_g}} & 0 & \frac{1}{\sqrt{1+C_g}}
    \end{pmatrix}
    \begin{pmatrix}
    1 & 0 & 0 & 0\\
    0 & 1 & 0 & 0\\
    0 & 0 & r & 0 \\
    0 & 0 & 0 & 1/r
    \end{pmatrix}
    \begin{pmatrix}
    0 & \frac{1}{\sqrt{1+C_g}} & \frac{-\sqrt{C_g}}{\sqrt{1+C_g}} & 0 \\
    \frac{1}{-\sqrt{1+C_g}} & 0 & 0 & \frac{-\sqrt{C_g}}{\sqrt{1+C_g}} \\
    0 & \frac{\sqrt{C_g}}{\sqrt{1+C_g}} & \frac{1}{\sqrt{1+C_g}} & 0 \\
    \frac{-\sqrt{C_g}}{\sqrt{1+C_g}} & 0 & 0 & \frac{1}{\sqrt{1+C_g}}
    \end{pmatrix},
\end{equation}
where we denoted the local effective squeezing
$r=\frac{1+C_g+2\sqrt{C_\nu}}{1+C_g-2\sqrt{C_\nu}}$. Obviously, the decomposition gives two beam-splitters sandwiched by a single mode amplifier, and the microwave squeezing parameter $C_\nu$ only appears in the $\mathbf{D}$ matrix. If we look at the quadrature transformations from optical $\hat{a}$ to microwave $\hat{b}$, it has the following structure
\begin{equation}
    \begin{split}
        \hat{x}^b_\mathrm{out,c}=&\frac{\sqrt{C_g}(1+r)}{1+C_g}\hat{p}^a_\mathrm{in,c}+\frac{{r-C_g}}{(1+C_g)}\hat{x}^b_\mathrm{in,c},\\
        \hat{p}^b_\mathrm{out,c}=&-\frac{\sqrt{C_g}(1+{1/r})}{1+C_g}\hat{x}^a_\mathrm{in,c}+\frac{{1/r}-C_g}{1+C_g}\hat{p}^b_\mathrm{in,c}.
    \end{split}
\end{equation}
We see the reflection can be canceled by tuning the squeezing, e.g., to satisfy $r=C_g$ or $1/r=C_g$, which gives the condition $C_\nu=\frac{1}{4}(1+C_g)^2$. Clearly, the introduced squeezing can make one quadrature impedance matched while the other is not, reducing the transducer to the class $[2,1]$ and a perfect transducer can be obtained, with a divergent quantum capacity.

\color{black}

\section{D---The Bogoliubov transformation}

In this section, we give the derivation of the Bogoliubov transformation used in the main text. Let's  consider the dynamics of the microwave squeezing
\begin{equation}
    \hat{H}_\nu=\hbar\Delta_\text{e}\hat{b}^\dagger\hat{b}+\hbar\nu(e^{-i\theta}\hat{b}^{\dagger^2}+e^{i\theta}\hat{b}^2).
\end{equation}
Including the dissipation, the mode dynamics follows the Langevin equation $d\mathbf{V}/dt=\mathbf{A}\mathbf{V}$, expanded as
\begin{equation}
    \frac{d}{dt}
    \begin{pmatrix}
    \hat{b}\\ \hat{b}^\dagger
    \end{pmatrix}=
    \begin{pmatrix}
    -i\Delta_\text{e}-\frac{\kappa_\text{e}}{2} & -i 2\nu e^{-i\theta} \\
    i 2\nu e^{i\theta} & i\Delta_\text{e}-\frac{\kappa_\text{e}}{2}
    \end{pmatrix}
    \begin{pmatrix}
    \hat{b}\\ \hat{b}^\dagger
    \end{pmatrix}.
\end{equation}
The right hand side defines a non-Hermitian dynamical matrix $\mathbf{A}$, which can be diagonalized according to $\mathbf{A}=\mathbf{W^{-1}DW}$ with the transformation matrix
\begin{equation}
    \mathbf{W}=
    \begin{pmatrix}
    \cosh(r) & e^{-i\theta}\sinh(r)\\
    e^{i\theta}\sinh(r) & \cosh(r)
    \end{pmatrix}
\end{equation}
and the corresponding eigenvalue matrix $\mathbf{D}=\text{Diag}[
-\frac{\kappa_\text{e}}{2}-i\sqrt{\Delta_\text{e}^2-(2\nu)^2}, 
-\frac{\kappa_\text{e}}{2}+i\sqrt{\Delta_\text{e}^2-(2\nu)^2}]$. The parameter $r$ is the \textit{effective squeezing} used in the transformation which satisfies $\tanh(2r)=2\nu/\Delta_\text{e}$ (note this effective squeezing is different from the microwave mode squeezing, although related). {It is worth mentioning that we need $\Delta_\text{e}>2\nu$ to guarantee a unitary Bogoliubov transformation. Practically, the detuning can be chosen to to smaller than the squeezing, although a Bogoliubov transformation can still be defined, the transformation is generally not unitary \cite{balian1969}, and identifying the corresponding quantum channel is an interesting topic for the future.} Define $\beta\equiv 2\nu/\Delta_\text{e}$. {When $\beta\sim 1$, the system gets to the {amplification threshold}. The effective squeezing factor $r$ quickly increases near the threshold ($\beta\sim1$) which as shown in the main text is the condition of boosting the transduction performance. } 

The diagonalization enables us to define a new eigen-mode $\hat{b}_\text{s}$ through the transformation
\begin{equation}
    \hat{b}_\text{s}=\cosh(r)\hat{b}+e^{-i\theta}\sinh(r)\hat{b}^\dagger
\end{equation}
with the corresponding mode frequency $\omega_\text{s}=\sqrt{\Delta_\text{e}^2-(2\nu)^2}$, and the mode dissipation rate stays the same $\kappa_\text{s}=\kappa_\text{e}$. This new mode is called the Bogoliubov squeezing mode as we used in the main text. The system total Hamiltonian can be written in the following form in terms of the Bogoliubov mode
\begin{equation}
    \begin{split}
        \hat{H}_\text{s}/\hbar=&-\Delta_\text{o}\hat{a}^\dagger\hat{a}+\omega_\text{s}\hat{b}_\text{s}^\dagger\hat{b}_\text{s}+g_\text{s}(\hat{a}^\dagger\hat{b}_\text{s}+\hat{a}\hat{b}^\dagger_\text{s})
    \end{split}
\end{equation}
which is the expression shown in the main text with the rotating wave approximation applied.

\section{E---Squeezing induced noise elimination}

The enhancement of coupling strength comes with a cost---the Bogoliubov mode will see a noise bath amplified by the squeezing. This noise could destroy all the quantum feature of the transduction channel. We write the squeezing-amplified noise operator as $\hat{b}_\nu=\cosh(r)\hat{b}_\mathrm{th}+e^{-i\theta}\sinh(r)\hat{b}_\mathrm{th}^\dagger$, where $\hat{b}_\mathrm{th}$ and $\hat{b}^\dagger_\mathrm{th}$ are the noise operators of microwave thermal bath with thermal photon $n_\mathrm{th}$. For vacuum bath $n_\mathrm{th}=0$, which is assumed in the main text (Note for thermal temperature $T\sim 1$ mK and the microwave frequency $\omega_\mathrm{e}\sim 10$ GHz, the thermal photon $n_\mathrm{th}\sim 10^{-200}$, which is negligible). Obviously, the noise photon of the squeezed bath is $n_\nu=\cosh(2r)n_\mathrm{th}+\sinh^2(r)$, as shown in the main text.

To suppress this noise, we may input a broadband squeezed vacuum to the microwave port with squeezing factor $\lambda$. Effectively, the squeezed-vacuum field can be considered as the thermal bath couple to the microwave mode with thermal photon $n_\mathrm{th}=\sinh^2\lambda$. The squeezed noise operator $\hat{b}_\nu$ should be written in the following form
\begin{equation}
\begin{split}
    \hat{b}_\nu=& \cosh(r)[\cosh(\lambda)\hat{b}_\mathrm{th}+e^{-i\phi}\sinh(\lambda)\hat{b}^\dagger_\mathrm{th}]+e^{-i\theta}\sinh(r)[\cosh(\lambda)\hat{b}^\dagger_\mathrm{th}+e^{i\phi}\sinh(\lambda)\hat{b}_\mathrm{th}]\\
    =& [\cosh(r)\cosh(\lambda)+e^{-i(\theta-\phi)}\sinh(r)\sinh(\lambda)]\hat{b}_\mathrm{th}+[e^{-i\phi}\cosh(r)\sinh(\lambda)+e^{-i\theta}\sinh(r)\cosh(\lambda)]\hat{b}^\dagger_\mathrm{th}.
\end{split}    
\end{equation}
Note $\hat{b}_\mathrm{th}$ is still the vacuum or thermal noise operator that the microwave mode originally couples to. This means the Bogoliubov mode will see a thermal reservoir with thermal photon (assuming $n_\mathrm{th}=0$)
\begin{equation}
\begin{split}
    n_\nu=&[e^{i\phi}\cosh(r)\sinh(\lambda)+e^{i\theta}\sinh(r)\cosh(\lambda)][e^{-i\phi}\cosh(r)\sinh(\lambda)+e^{-i\theta}\sinh(r)\cosh(\lambda)]\\
    =&\cosh^2(r)\sinh^2(\lambda)+\sinh^2(r)\cosh^2(\lambda)+\frac{1}{2}\cos(\theta-\phi)\sinh(2r)\sinh(2\lambda).
\end{split}    
\end{equation}
Obviously, if we pick $r=\lambda$ and $\theta-\phi=\pm k\pi$ with $k=1,3,5,...$, we can get $n_\nu=0$. It means the squeezed noise can be totally eliminated. The Bogoliubov mode is effectively couples to a vacuum reservoir. This effect can be understood as the phase matching: the original vacuum bath is first squeezed along the direction $\phi/2$ with squeezing factor $\lambda$, then this squeezing is totally cancelled by Bogoliubov mode squeezing along the direction $\theta/2=\phi/2\pm k\pi/2$ with squeezing factor $r=\lambda$. In the phase space, the elongated noise distribution becomes a circle again as seen by the Bogoliubov mode.

Figure~\ref{fig5} shows more data about the capacity enhancement with and without squeezing amplified noise elimination. The Fig.~\ref{fig5}(b) and (c) are shown in the main text. Note the data becomes inaccurate when $\beta$ really approaches one because the Bogoliubov framework will fail in the regime, as discussed in the main text.

\begin{figure}[t]
\centering
\includegraphics[width=\columnwidth]{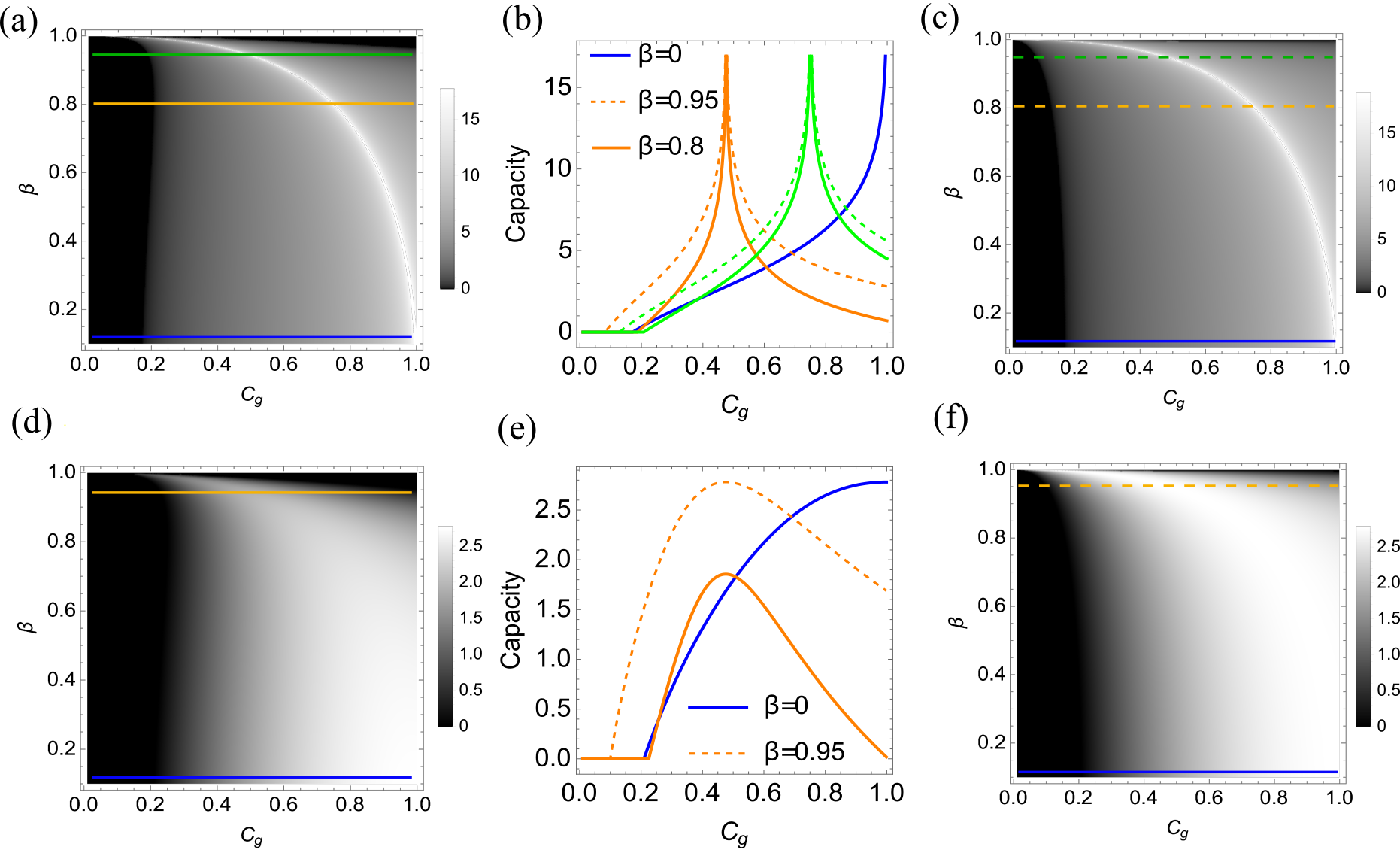}
\caption{The capacity lower bound by scanning the parameter $C_g$ and $\beta$. The unit extraction ratios are use in (a),(b) and (c), while $\zeta_\mathrm{e}=0.97, \zeta_\mathrm{o}=0.9$ for (d), (e) and (f). Noise eliminations are done in (c) and (f). The curves in (b) and (e) trace the solid and dashed line in (a)(c) and (d)(f), correspondingly. \label{fig5}}
\end{figure}

\end{document}